\documentclass[twocolumn]{aastex631}

\newcommand{\eqref}[1]{(\ref{#1})}

\shorttitle{HIP\,41378\,$\rm f$  }
\shortauthors{Alam et al.}

\begin{document}

\title{The First Near-Infrared Transmission Spectrum of HIP\,41378\,f, \\ a Low-Mass Temperate Jovian World in a Multi-Planet System}

\author[0000-0003-4157-832X]{Munazza K. Alam}
\affiliation{Carnegie Earth \& Planets Laboratory, 5241 Broad Branch Rd NW, Washington, DC 20015, USA}
\affiliation{Center for Astrophysics $|$ Harvard \& Smithsonian, 60 Garden Street, Cambridge, MA 02138, USA}

\author[0000-0002-4207-6615]{James Kirk}
\affiliation{Center for Astrophysics $|$ Harvard \& Smithsonian, 60 Garden Street, Cambridge, MA 02138, USA}

\author[0000-0001-8189-0233]{Courtney D. Dressing}
\affiliation{Astronomy Department, 501 Campbell Hall $\#$3411, University of California, Berkeley, CA 94720, USA}

\author[0000-0003-3204-8183]{Mercedes L\'{o}pez-Morales}
\affiliation{Center for Astrophysics $|$ Harvard \& Smithsonian, 60 Garden Street, Cambridge, MA 02138, USA}

\author[0000-0003-3290-6758]{Kazumasa Ohno}
\affiliation{Department of Astronomy and Astrophysics, University of California Santa Cruz, 1156 High St, Santa Cruz, CA 95064, USA}

\author[0000-0002-8518-9601]{Peter Gao}
\affiliation{Carnegie Earth \& Planets Laboratory, 5241 Broad Branch Rd NW, Washington, DC 20015, USA}

\author[0000-0001-6519-1598]{Babatunde Akinsanmi}
\affiliation{Instituto de Astrof\'isica e Ci\^encias do Espa\c{c}o, Universidade do Porto, CAUP, Rua das Estrelas, PT4150-762 Porto, Portugal}
\affiliation{Departamento de Fisica e Astronomia, Faculdade de Ciencias, Universidade do Porto, Rua Campo Alegre, 4169-007 Porto, Portugal}
\affiliation{Observatoire Astronomique de l'Université de Genève, Chemin Pegasi 51, Versoix, Switzerland}

\author[0000-0002-3586-1316]{Alexandre Santerne}
\affiliation{Aix Marseille University, CNRS, CNES, LAM, Marseille, France}

\author[0000-0002-2805-5869]{Salom\'{e} Grouffal}
\affiliation{Aix Marseille University, CNRS, CNES, LAM, Marseille, France}

\author[0000-0002-0601-6199]{Vardan Adibekyan}
\affiliation{Instituto de Astrof\'isica e Ci\^encias do Espa\c{c}o, Universidade do Porto, CAUP, Rua das Estrelas, PT4150-762 Porto, Portugal}
\affiliation{Departamento de Fisica e Astronomia, Faculdade de Ciencias, Universidade do Porto, Rua Campo Alegre, 4169-007 Porto, Portugal}

\author[0000-0003-2434-3625]{Susana C. C. Barros}
\affiliation{Instituto de Astrof\'isica e Ci\^encias do Espa\c{c}o, Universidade do Porto, CAUP, Rua das Estrelas, PT4150-762 Porto, Portugal}
\affiliation{Departamento de Fisica e Astronomia, Faculdade de Ciencias, Universidade do Porto, Rua Campo Alegre, 4169-007 Porto, Portugal}

\author[0000-0003-1605-5666]{Lars A. Buchhave}
\affiliation{DTU Space, National Space Institute, Technical University of Denmark, Elektrovej 328, DK-2800 Kgs. Lyngby, Denmark}

\author[0000-0002-1835-1891]{Ian J.\ M.\ Crossfield}
\affiliation{Department of Physics and Astronomy, University of Kansas, Lawrence, KS, 66045, USA}

\author[0000-0002-8958-0683]{Fei Dai} 
\affiliation{Division of Geological and Planetary Sciences,
1200 E California Blvd, Pasadena, CA, 91125, USA}

\author[0000-0001-6036-0225]{Magali Deleuil}
\affiliation{Aix Marseille University, CNRS, CNES, LAM, Marseille, France}

\author[0000-0002-8965-3969]{Steven Giacalone}
\affil{Department of Astronomy, University of California Berkeley, Berkeley, CA 94720, USA}

\author[0000-0003-3742-1987]{Jorge Lillo-Box}
\affiliation{Centro de Astrobiolog\'ia (CAB, CSIC-INTA), Depto. de Astrof\'isica, ESAC campus, 28692, Villanueva de la Ca\~nada (Madrid), Spain}

\author[0000-0002-5251-2943]{Mark Marley}
\affiliation{Lunar \& Planetary Laboratory, University of Arizona, Tucson, AZ, 85721, USA}

\author[0000-0002-7216-2135]{Andrew W. Mayo}
\affiliation{Astronomy Department, 501 Campbell Hall $\#$3411, University of California, Berkeley, CA 94720, USA}
\affiliation{Centre for Star and HIP\,41378\,$\rm f$ormation, Natural History Museum of Denmark \& Niels Bohr Institute, University of Copenhagen, \O ster Voldgade 5-7, DK-1350 Copenhagen K., Denmark}

\author[0000-0001-7254-4363]{Annelies Mortier}
\affiliation{KICC \& Cavendish Laboratory, University of Cambridge, J.J. Thomson Avenue, Cambridge CB3 0HE, UK}

\author[0000-0003-4422-2919]{Nuno C. Santos}
\affiliation{Instituto de Astrof\'isica e Ci\^encias do Espa\c{c}o, Universidade do Porto, CAUP, Rua das Estrelas, PT4150-762 Porto, Portugal}
\affiliation{Departamento de Fisica e Astronomia, Faculdade de Ciencias, Universidade do Porto, Rua Campo Alegre, 4169-007 Porto, Portugal}

\author[0000-0001-9047-2965]{S\'ergio G. Sousa}
\affiliation{Instituto de Astrof\'isica e Ci\^encias do Espa\c{c}o, Universidade do Porto, CAUP, Rua das Estrelas, PT4150-762 Porto, Portugal}

\author[0000-0002-1845-2617]{Emma V. Turtelboom}
\affiliation{Astronomy Department, 501 Campbell Hall $\#$3411, University of California, Berkeley, CA 94720, USA}

\author[0000-0003-1452-2240]{Peter J.\ Wheatley}
\affiliation{Department of Physics, University of Warwick, Gibbet Hill Road, Coventry CV4 7AL, UK}

\author[0000-0001-7246-5438]{Andrew M. Vanderburg}
\affiliation{Massachusetts Institute of Technology, 77 Massachusetts Avenue, 37-241, Cambridge, MA 02139, USA}

\begin{abstract} 

We present a near-infrared transmission spectrum of the long period (P=542 days), temperate ($\rm T_{eq}$=294 K) giant planet HIP\,41378\,$\rm f$   obtained with the Wide-Field Camera 3 (WFC3) instrument aboard the \textit{Hubble Space Telescope} (HST). With a measured mass of $12 \pm 3$\,$M_{\earth}$ and a radius of $9.2 \pm 0.1$\,$R_{\earth}$, HIP\,41378\,$\rm f$   has an extremely low bulk density ($0.09 \pm 0.02 \,  {\rm g} \, {\rm cm}^{-3}$). We measure the transit depth with a median precision of 84 ppm in 30 spectrophotometric channels with uniformly-sized widths of 0.018 $\mu$m. Within this level of precision, the spectrum shows no evidence of absorption from gaseous molecular features between 1.1--1.7 $\mu$m. Comparing the observed transmission spectrum to a suite of 1D radiative-convective-thermochemical-equilibrium forward models, we rule out clear, low-metallicity atmospheres and find that the data prefer high-metallicity atmospheres or models with an additional opacity source, such as high-altitude hazes and/or circumplanetary rings. We explore the ringed scenario for HIP\,41378\,$\rm f$ further by jointly fitting the \textit{K2} and HST light curves to constrain the properties of putative rings. We also assess the possibility of distinguishing between hazy, ringed, and high-metallicity scenarios at longer wavelengths with JWST. HIP\,41378\,$\rm f$ provides a rare opportunity to probe the atmospheric composition of a cool giant planet spanning the gap in temperature, orbital separation, and stellar irradiation between the Solar System giants, directly imaged planets, and the highly-irradiated hot Jupiters traditionally studied via transit spectroscopy. 
\end{abstract}


\section{Introduction} 
\label{sec:intro}

The Solar System gas and ice giants host ring systems, although the origins of these rings remain mysterious (e.g., \citealt{DePater17,DePater18,Hedman21}). Whereas Saturn's massive rings are rich in water ice (\citealt{Cuzzi1998,poulet_et_al2003,Nicholson2005}), the less massive rings of Uranus and Neptune have a higher content of rocky particles (\citealt{Tiscareno2013}) and Jupiter's tenuous rings are composed of micron-sized dust particles (\citealt{DePater17}). Massive ring systems like that of Saturn's may form from collisions (e.g., \citealt{Pollack75}), or tidal disruptions of primordial satellites (\citealt{Canup10}) or passing objects (e.g., \citealt{Dones91}). Despite the prevalence of rings around the Solar System giants, exoplanet characterization efforts have not yet yielded conclusive observational evidence of circumplanetary rings (\citealt{Heising15,Aizawa18}). The presence of ring systems around giant exoplanets, however, may explain planets with large measured radii and unusually low bulk densities.  

HIP~41378 is a nearby bright  ($V = 8.93$), late F-type star hosting five transiting planets (\citealt{vanderburg_et_al2016,berardo_et_al2019}). With $\rm T_{eq}$=294 K (assuming full heat redistribution and zero Bond albedo), the outermost HIP\,41378\,$\rm f$   is an intriguing target for atmospheric characterization because it is significantly colder than the giant exoplanets typically probed by ground-based and space-based observations. It therefore provides a bridge between the highly-irradiated hot Jupiters studied via transmission spectroscopy; the young, wide-orbit, and massive giant planets or substellar objects studied via direct imaging;  and the colder, mature gas giants in the Solar System. With a measured mass of 12$\pm$3\,$M_{\earth}$ and a radius of 9.2$\pm$0.1\,$R_{\earth}$ \citep{santerne_et_al2019}, HIP\,41378\,$\rm f$ stands out as one of the lowest bulk density planets discovered to date ($0.09\pm0.02 \,  {\rm g} \, {\rm cm}^{-3}$). 

Rings have been shown theoretically to inflate an exoplanet's radius measured through transits, thereby decreasing the inferred bulk density 
\citep{piro+vissapragada2020, akinsanmi_et_al2020}. As discussed in \citet{akinsanmi_et_al2020}, the ring-induced transit depth enhancement is expected to be chromatic: deeper transits occur at wavelengths where the ring is optically thick. Based on detailed modeling of the observed \textit{K2} photometry, the most likely ring scenario for HIP\,41378\,$\rm f$   is a Uranian-like bulk density of 1.23\,${\rm g} \, {\rm cm}^{-3}$ and a ring extending from 1.05 to 2.59 times the planetary radius inclined at an angle of $\sim$25$^\circ$ from the sky plane \citep{akinsanmi_et_al2020}. While HIP\,41378\,$\rm f$   is too close to its host star to support icy rings, the planet could instead be orbited by rings composed of small, porous rocky particles \citep{piro+vissapragada2020, akinsanmi_et_al2020}. In this ringed model, the planet radius would be $\rm R_p = 3.7 \pm 0.3 \, \rm R_\oplus$ (compared to $\rm R_p = 9.21 \pm 0.01 \, \rm R_\oplus$ for the ringless case).

Alternatively, if HIP\,41378\,$\rm f$   does not possess rings, then it may be a member of the rare class of ``super-puff" exoplanets, which have been inferred to possess gas mass fractions far greater than the more common mini-Neptunes with similiar masses ($>$10\% vs.~a few \%, respectively; \citealt{Lopez14}). Super-puffs have been hypothesized to form in a less opaque region of the protoplanetary disk \citep{lee+chiang2016}, followed by inward migration to account for their large gas mass fractions. However, despite their low densities and correspondingly large atmospheric scale heights, all transmission spectra of super-puffs observed to date have been flat (e.g., \citealt{libby-roberts_et_al2020,chachan_et_al2020}), suggesting that flat spectra are a general property of the population. High-altitude photochemical hazes have been considered to explain both the flat spectra and the large radii of super-puffs (\citealt{gao+zhang2020,Ohno2021}), although dusty outflows driven by expected high mass loss rates \citep{wang+dai19} may also explain their effective radii. 

Here we present the near-infrared transmission spectrum of HIP\,41378\,$\rm f$   obtained with the \emph{Hubble Space Telescope} Wide-Field Camera 3 (HST/WFC3), which we use to constrain the planet's atmospheric composition and explore the presence of circumplanetary rings. In Section \ref{sec:obs_dr}, we describe our HST observations and data reduction procedure. Section \ref{sec:lc_fitting} details our methods for fitting the transit light curves. In Section \ref{sec:results}, we present the near-infrared transmission spectrum of HIP\,41378\,$\rm f$. In Section \ref{sec:discussion}, we interpret our results using atmospheric models, explore the possibility of rings, and compare HIP\,41378\,$\rm f$   to other planets with similar masses and radii. Section \ref{sec:conclusions} summarizes our results.

\section{Observations \& Data Reduction} 
\label{sec:obs_dr}
\subsection{Observations}
\label{sec:observations}

We observed a single primary transit of HIP\,41378\,$\rm f$   with HST/WFC3 using the G141 grism, which provides spectroscopy between 1.125-1.643 $\mu$m at a spectral resolving power of R$\sim$130 around $\lambda$=1.4 $\mu$m. Taken as part of GO 16267 (PI: Dressing) on UT 2021 May 19-21, the observations were scheduled over three consecutive HST visits of six orbits each to accommodate the target's long transit duration (18.998 hours; \citealt{vanderburg_et_al2016}). To ensure that the target remained centered in the instrument's field of view, we took an image of the target using the F126N filter with an exposure time of 7.317 seconds at the beginning of the first and third visits as well as at the beginning of the last orbit of the second visit. We then obtained time series spectroscopy with the G141 grism, and used round-trip spatial scanning for all visits with a scan rate of 0.419 arcsec~s$^{-1}$ to permit taking longer exposures without saturating the detector \citep{McCullough12}. 

Due to South Atlantic Anomaly (SAA) passages during the transit, we varied the sampling sequence for affected orbits. The first three were impacted by SAA passages, making roughly half, a third, and a quarter of the respective orbits unusable. The fourth through tenth orbits were less affected, so we were able to use almost all exposures taken during these orbits. For the remaining orbits (orbits 11-18), we again faced interruptions due to the SAA. Ultimately, we used the SPARS10 sampling sequence with 7-9 non-destructive reads per exposure (NSAMP = 7-9). As a result of these NSAMP changes, the total integration times ranged from 37.010 to 51.703 seconds and scans were taken across approximately 126-178 pixel rows in the cross-dispersion direction. With this instrument setup, we read out a 512$\times$512 pixel subarray for each science exposure and obtained a total of 274 science exposures over the 18 orbits observed. 

\vspace{-0.2cm}

\subsection{Data Reduction}
\label{sec:data_reduction}

We reduced the observations for this program using the methods outlined in \citet{Alam20}, which we briefly summarize here. We started our analysis using the bias-corrected, flat-fielded {\tt ima} images from the CALWF3 pipeline. The flux for each exposure was extracted by taking the difference between successive reads and then performing a background subtraction to suppress contamination from nearby stars. For the background subtraction, we subtracted the median flux from a box 32 pixels away from the spectrum. To correct for cosmic ray events, we followed the procedure of \citet{Nikolov14}.  

Next, we extracted stellar spectra by summing the flux within a rectangular aperture. To determine the size of the aperture (accounting for the different-sized spatial scans; see Section \ref{sec:observations}), we fit for the aperture width along the dispersion and cross-dispersion axes. We scanned each row of the raw {\tt ima} images and fit a top-hat function to the data to determine the center of the PSF (i.e., the center of the 2D spectrum along the cross-dispersion axis). To determine the center of the spectrum along the dispersion direction, we scanned each column and fit a top-hat distribution to the data. We then used these fitted x and y center points as initial guesses for calculating the centroid positions on each image using the flux-weighted first moments in x and y pixel position. To determine the wavelength solution, we cross-correlated each stellar spectrum to a grid of model spectra from the WFC3 Exposure Time Calculator (ETC) with temperatures ranging from 4060-9230~K. To determine shifts along the dispersion axis, we used the closest matching model of 6200~K. The final wavelengths assigned to each element of the spectrum are the wavelengths from the ETC model with the shift applied.
\vspace{-0.2cm}
\section{Light Curve Fits}
\label{sec:lc_fitting}

\begin{figure*}
    \centering
    \includegraphics[scale=0.95]{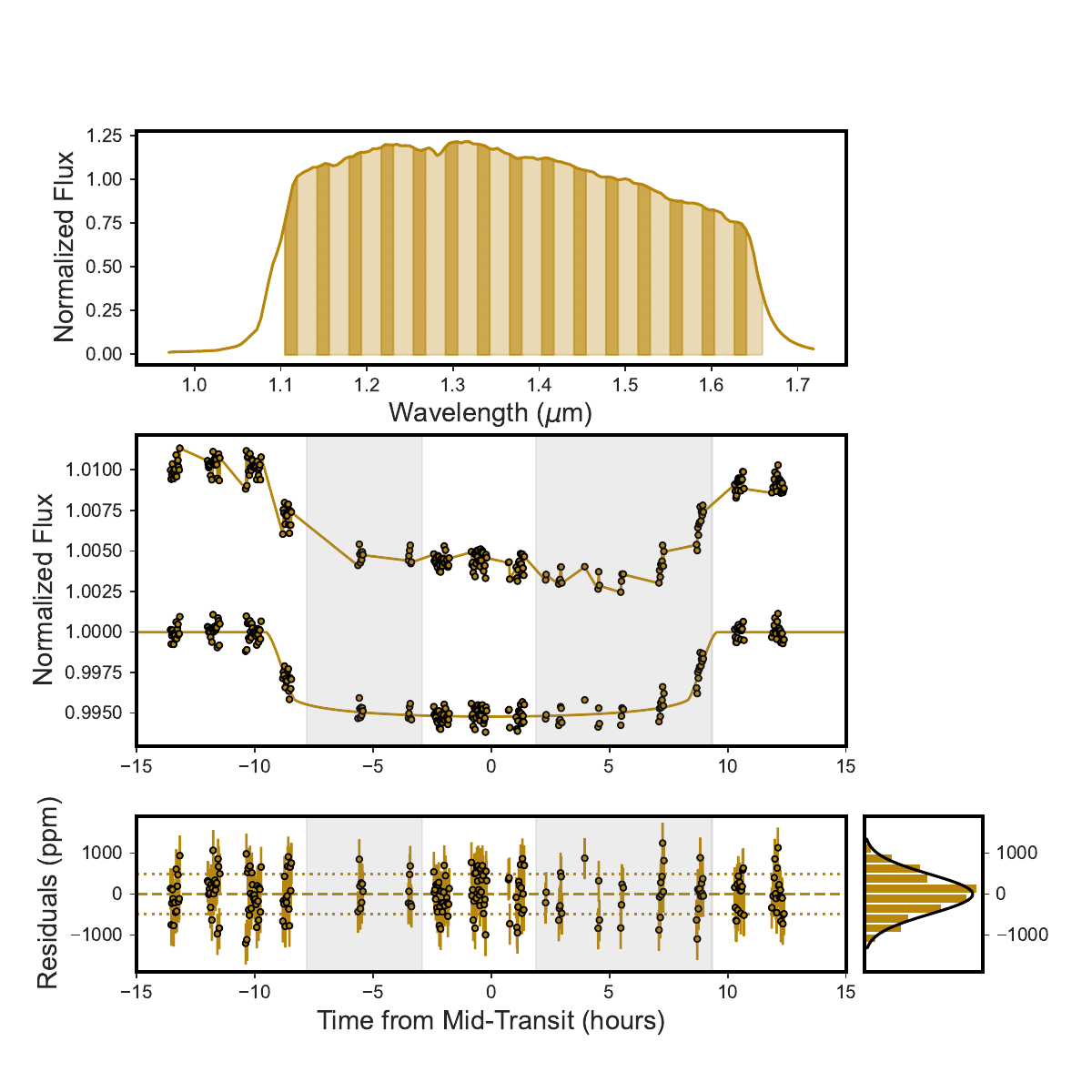}
    \caption{Top: Example HIP\,41378\,$\rm f$   stellar spectrum for the HST/WFC3 G141 grism. The vertical bands denote the 0.018 $\mu$m wavelength channels adopted for the spectroscopic light curves. Middle: Raw (top) and detrended (bottom) white light curve, excluding the first orbit and the first exposure of each subsequent orbit (points). The raw light curve has been shifted vertically by an arbitrary constant for clarity. Overplotted is the best-fitting analytic light curve model (line). Epochs most affected by South Atlantic Anomaly passages are denoted by the gray shading. Bottom: RMS residuals of the transit fit in ppm (left; points) and 1-$\sigma$ range of the residuals (dotted lines), as well as the the distribution of residuals (right).}
    \label{fig:wlc_lm}
\end{figure*}

To extract the 1.1-1.7 $\mu$m transmission spectrum, we fit the transit light curves using the fitting routine detailed in \citet{Kirk17,Kirk18,Kirk19,Kirk21} and \citet{Alam21}. Briefly, we modeled the analytic transit light curves \citep{Mandel02} using {\tt Batman} \citep{batman} and implemented a Gaussian process (GP) with the {\tt george} code \citep{Ambikasaran14} to model noise in the data. We fixed the non-linear limb darkening coefficients to the theoretical values from 3D stellar models \citep{Magic15}, assuming stellar properties consistent with \citet{santerne_et_al2019} ($\rm T_{eff}$=6320\,K, log(g)=4.294, [Fe/H]=-0.1).

To fit the white light curve (1.104-1.661\,$\mu$m), we fixed the system parameters to those in \citealt{santerne_et_al2019} ($\rm P$=542.07975 days, $\rm a/R_\star$=231.417, $i$=89.971$^{\circ}$, $e$=0). We fit for the time of mid-transit $\rm T_{0}$, the scaled planetary radius $\rm R_p/R_\star$, and the GP hyperparameters. For the white light curve, we used three squared-exponential GP kernels with the orbital phase of HST, the wavelength shift in the stellar spectra, and time as the three input vectors -- after standardizing each (subtracting the mean and dividing by the standard deviation to put each on a common scale). We fit for the natural logarithm of the inverse length-scale for each kernel, in addition to the amplitude of the GP and a white noise term. The GP was therefore described by five free hyperparameters. We placed truncated uniform-in-log-space priors on the GP hyperparameters. The amplitude was bounded between 0.01 and $100\times$ the out-of-transit variance and the length scales were bounded between the minimum spacing and $3\times$ the maximum spacing of the standardized input vectors. The white noise term was bounded between 0.1 and 1000\,ppm.

To sample the parameter space for the white light curve and spectroscopic light curves, we ran a Markov Chain Monte Carlo (MCMC) using {\tt emcee} after clipping $>$4-$\sigma$ outliers from a running median computed for each light curve, which clipped 0--2 points per light curve. We optimized the GP hyperparameters to the out-of-transit data to find the starting locations for the hyperparameters. We followed the example in the \texttt{george} documentation\footnote{\url{https://george.readthedocs.io/en/latest/tutorials/hyper/}} and used \texttt{scipy.optimize} \citep{scipy} and a `L-BFGS-B' algorithm to perform the optimization of the GP hyperparameters. The starting value for $\rm R_{p}/R_{\star}$ was taken to be 0.0663 \citep{santerne_et_al2019} and the starting $\rm T_{0}$ value of BJD 2459354.6 was determined by visual inspection of the light curve. We initialized the chains around these values and ran the MCMC with 210 walkers for a 2000 step burn-in, followed by a 6000 step chain for our posterior and parameter estimates. The number of samples was $72\times$ the autocorrelation length, greater than the $50\times$ autocorrelation length which in general indicates convergence in \texttt{emcee}\footnote{\url{https://dfm.io/posts/autocorr/}}. The best-fit white light curve is shown in Figure \ref{fig:wlc_lm}, with $\rm R_p/R_\star$ = 0.068602 $^{+0.002684} _{-0.003370}$ and $\rm T_{0}$ = BJD 2459355.101374 $^{+0.001919} _{-0.001888}$.

For the binned light curve fits, we used a common mode correction. This correction involved removing the best-fitting white light systematics model and the residuals to the white light fit from each binned light curve prior to fitting (e.g., \citealt{Gibson14,Alam20}). As a result, we were able to use a simpler systematics model for the binned light curves and therefore only used two GP kernels (HST phase and wavelength shift). We also held $\rm T_0$ fixed to the best-fit value from the white light curve fit, resulting in five free parameters per binned light curve ($\rm R_p/R_\star$ and the four GP hyperparameters). We then proceeded with the MCMC as for the white light curve fit. The measured $\rm R_{p}/R_{\star}$ values for each spectroscopic light curve\footnote{The spectroscopic light curves are shown \href{https://doi.org/10.6084/m9.figshare.17373572}{here} in the online supplementary material.} are presented in Table \ref{tab:tr_spec}. 

\section{Results}
\label{sec:results}

\begin{figure*}
    \centering
    \includegraphics[width=\textwidth, trim={2.1cm 2.0cm 1.9cm 2.5cm}]{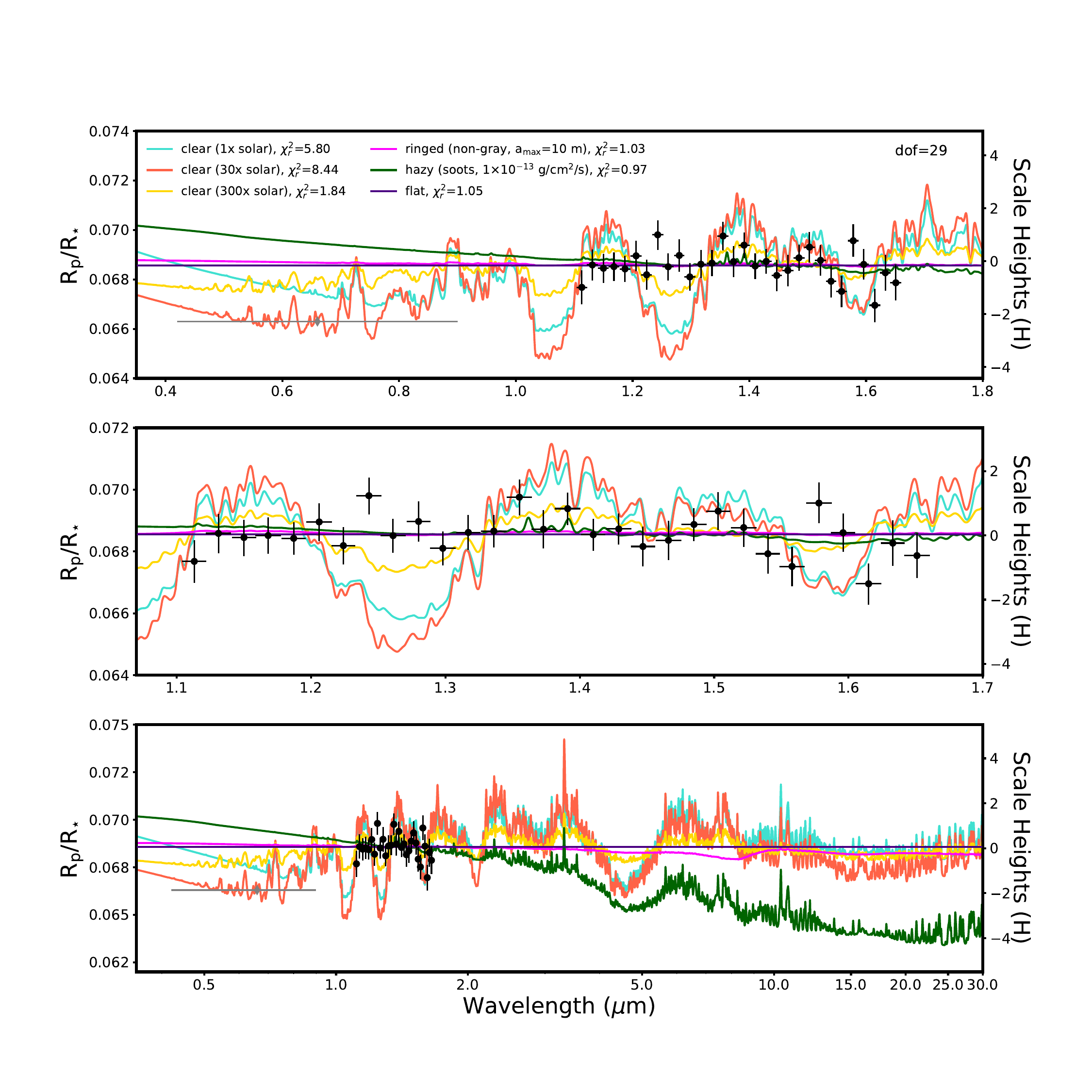}
    \caption{Broadband transmission spectrum for HIP\,41378\,$\rm f$ (black points), compared to 1D radiative-convective forward models (colored lines) for a clear atmosphere with metallicities 1$\times$, 30$\times$, and 300$\times$ solar, example hazy and ringed models, as well as a flat line (gray atmosphere). Top: The WFC3 transmission spectrum and the measured optical transit depth measurement from \textit{K2} (gray diamond; \citealt{santerne_et_al2019}). Middle: Zoom-in to the WFC3 data presented in this work. Bottom: Zoom-out to the full spectrum, including mid-infrared wavelengths accessible with JWST.}
    \label{fig:hip41378_trspec}
\end{figure*}

The near-infrared (1.1-1.7 $\mu$m) transmission spectrum of HIP\,41378\,$\rm f$   is shown in Figure \ref{fig:hip41378_trspec}. The $\rm R_p/R_\star$ values measured from our WFC3 transmission spectrum (Table~\ref{tab:tr_spec}) vary between 0.067 and 0.070, consistent within 2-$\sigma$ with the optical transit depths measured from the two \textit{K2} observations\footnote{We note that the WFC3 white light curve has errorbars $\sim$30 times larger than the \textit{K2} observations, and the lower limit of the WFC3 white light curve is consistent within  1-$\sigma$ with the \textit{K2} $\rm R_{p}/R_{\star}$ value derived in \citet{santerne_et_al2019}.} 
(0.0672$\pm$0.0013, \citealt{vanderburg_et_al2016}; $0.06602^{+0.00017}_{-0.00016}$, \citealt{berardo_et_al2019}; 0.0663$\pm$0.0001, \citealt{santerne_et_al2019}). Within the precision of our observations, we find that the spectrum does not display any large absorption features from gaseous molecules, with maximum deviations spanning $\sim$2 scale heights (i.e., $\sim1200$\,km, assuming a H/He-dominated atmospheric composition with a mean molecular weight of 2.3~amu). 

\vspace{-0.2cm}

\begin{deluxetable}{cccccc}
\tabletypesize{\scriptsize}
\tablewidth{0pt}
\tablecolumns{6}
\tablecaption{Broadband HST/WFC3 transmission spectrum for HIP~41378 f and adopted non-linear limb darkening coefficients 
\label{tab:tr_spec}}

\tablehead{\colhead{$\lambda$ ($\mu$m)}  & \colhead{$\rm R_{p}/R_{*}$} & \colhead{$c_{1}$}  & \colhead{$c_{2}$}  & \colhead{$c_{3}$}  & \colhead{$c_{4}$} }

\startdata 
1.104$-$1.122  &  0.067679 $^{+0.000692} _{-0.000693}$    & 0.6449  &  -0.2808  &  0.3552  &  -0.1548 \\
1.122$-$1.141  &  0.068585 $^{+0.000635} _{-0.000637}$    & 0.6534  &  -0.3110  &  0.3794  &  -0.1639 \\                            
1.141$-$1.159  &  0.068451 $^{+0.000580} _{-0.000598}$    & 0.6371  &  -0.2605  &  0.3191  &  -0.1396 \\                                      
1.159$-$1.178  &  0.068517 $^{+0.000584} _{-0.000587}$    & 0.6406  &  -0.2584  &  0.3035  &  -0.1340 \\                                      
1.178$-$1.196  &  0.068425 $^{+0.000524} _{-0.000528}$    & 0.6470  &  -0.2813  &  0.3155  &  -0.1335 \\                                      
1.196$-$1.215  &  0.068955 $^{+0.000606} _{-0.000617}$    & 0.6328  &  -0.2409  &  0.2684  &  -0.1154 \\                                      
1.215$-$1.233  &  0.068189 $^{+0.000606} _{-0.000605}$    & 0.6230  &  -0.1898  &  0.2075  &  -0.0945 \\                                      
1.233$-$1.252  &  0.069804 $^{+0.000593} _{-0.000607}$    & 0.6232  &  -0.1836  &  0.1922  &  -0.0884 \\                                      
1.252$-$1.271  &  0.068512 $^{+0.000536} _{-0.000550}$    & 0.6213  &  -0.1553  &  0.1457  &  -0.0694 \\                                      
1.271$-$1.289  &  0.068970 $^{+0.000650} _{-0.000659}$    & 0.6351  &  -0.1353  &  0.0720  &  -0.0408 \\                                      
1.289$-$1.308  &  0.068102 $^{+0.000553} _{-0.000549}$    & 0.6184  &  -0.1191  &  0.0842  &  -0.0426 \\                                      
1.308$-$1.326  &  0.068615 $^{+0.000569} _{-0.000561}$    & 0.6190  &  -0.1109  &  0.0633  &  -0.0320 \\                                      
1.326$-$1.345  &  0.068659 $^{+0.000529} _{-0.000532}$    & 0.6256  &  -0.1046  &  0.0418  &  -0.0232 \\                                      
1.345$-$1.364  &  0.069756 $^{+0.000577} _{-0.000587}$    & 0.6275  &  -0.0940  &  0.0124  &  -0.0081 \\                                      
1.364$-$1.382  &  0.068719 $^{+0.000621} _{-0.000618}$    & 0.6383  &  -0.0943  &- 0.0114  &   0.0046 \\                                      
1.382$-$1.401  &  0.069385 $^{+0.000522} _{-0.000526}$    & 0.6424  &  -0.0831  &- 0.0374  &   0.0160 \\                             
1.401$-$1.419  &  0.068542 $^{+0.000518} _{-0.000519}$    & 0.6488  &  -0.0783  &- 0.0556  &   0.0247 \\                                      
1.419$-$1.438  &  0.068728 $^{+0.000496} _{-0.000506}$    & 0.6495  &  -0.0572  &- 0.1034  &   0.0495 \\                                      
1.438$-$1.456  &  0.068161 $^{+0.000640} _{-0.000645}$    & 0.6716  &  -0.1004  &- 0.0728  &   0.0378 \\                                      
1.456$-$1.475  &  0.068362 $^{+0.000626} _{-0.000637}$    & 0.6829  &  -0.0861  &- 0.1108  &   0.0554 \\                                      
1.475$-$1.494  &  0.068873 $^{+0.000534} _{-0.000532}$    & 0.6897  &  -0.1098  &- 0.1023  &   0.0590 \\                                      
1.494$-$1.512  &  0.069306 $^{+0.000611} _{-0.000617}$    & 0.7115  &  -0.1676  &- 0.0578  &   0.0451 \\                                      
1.512$-$1.531  &  0.068772 $^{+0.000615} _{-0.000620}$    & 0.7538  &  -0.2127  &- 0.0480  &   0.0493 \\                                      
1.531$-$1.549  &  0.067925 $^{+0.000647} _{-0.000647}$    & 0.7838  &  -0.2858  &  0.0132  &   0.0297 \\                                      
1.549$-$1.568  &  0.067516 $^{+0.000636} _{-0.000646}$    & 0.7986  &  -0.3470  &  0.0823  &   0.0038 \\                                      
1.568$-$1.587  &  0.069565 $^{+0.000662} _{-0.000646}$    & 0.7781  &  -0.3858  &  0.1569  &  -0.0284 \\                                      
1.587$-$1.605  &  0.068607 $^{+0.000615} _{-0.000615}$    & 0.8379  &  -0.4884  &  0.2304  &  -0.0526 \\                                      
1.605$-$1.624  &  0.066955 $^{+0.000664} _{-0.000678}$    & 0.8540  &  -0.5049  &  0.2239  &  -0.0443 \\                                      
1.624$-$1.642  &  0.068270 $^{+0.000733} _{-0.000734}$    & 0.8801  &  -0.5741  &  0.2771  &  -0.0564 \\                                      
1.642$-$1.661  &  0.067865 $^{+0.000724} _{-0.000718}$    & 0.8508  &  -0.4979  &  0.2070  &  -0.0336 \\                                      
\enddata
\end{deluxetable}

\subsection{Comparisons to 1D Forward Models}

We compare our observed transmission spectrum to model spectra generated using a 1D radiative-convective-thermochemical equilibrium model \citep{saumon2008} assuming clear, solar-composition atmospheres with metallicities 1$\times$, 30$\times$, and 300$\times$ solar. 
Considering HIP\,41378\,$\rm f$'s low bulk density and featureless transmission spectrum, we also compare to models incorporating high-altitude hazes and circumplanetary rings. The hazy models were computed using the Community Aerosol and Radiation Model for Atmospheres \citep[CARMA;][]{Gao18} by adding a downward flux of haze particles to the 1$\times$ solar model atmosphere and tracking particle coagulation, sedimentation, and mixing (as in e.g., \citealt{Adams19}). We assumed spherical haze particles with compositions of soots and tholins for haze column production fluxes of $10^{-10}$, $10^{-11}$, $10^{-12}$, $10^{-13}$, and $10^{-14}$ $\rm g~\rm cm^{-2}~\rm s^{-1}$ (\citealt{lavvas2017,kawashima2019}). We considered one soot model and one tholin model at each of the five haze production rates for a total of 10 hazy models.   

We also computed transmission spectra with circumplanetary rings following the post-processing method described in a companion paper \citep{Ohno&Fortney22}. In short, the method computes the spectrum by summing the transmittance of the ring-free planetary disk and the circumplanetary ring outside of the planetary disk. Using Equation 3 of \citet{schlichting+chang2011}, we estimate a minimum ring particle size of $\sim$10~cm and a system age of 3.1$\pm$0.6~Gyr \citep{santerne_et_al2019}. Since the particle size is much larger than the relevant wavelength, we first assume a gray ring opacity. The gray rings model grid comprises 80 models assuming a solar composition atmosphere and an opaque ring with a morphology consistent with \citet{akinsanmi_et_al2020}. We varied the ring inclination between 21-28$^{\circ}$ from the sky plane (in increments of 1$^{\circ}$) and varied the inner ring radius between 1.02-1.11$\rm R_{0}$ (the ring-free transit radius of 0.35$\rm R_{J}$ from the clear 1$\times$ solar model, derived based on the measured mass from \citealt{santerne_et_al2019} and the inferred bulk density estimate from \citealt{akinsanmi_et_al2020}) in steps of 0.01. The outer ring radius was fixed to 2.55$\rm R_{0}$, the Roche radius beyond which ring particles would coagulate into a satellite. 

We also tested model grids for ring opacities computed using Mie theory and assuming a power-law size distribution for ring particles. The refractive indices are taken for astronomical silicates \citep{Draine03}.
We assumed a ring mass surface density of $100~{\rm g~{cm}^{-2}}$ with the largest particle size of $10\,{\rm m}$, similar to Saturnian rings. Since tiny particles might survive in optically thick rings \citep{schlichting+chang2011}, we set the smallest particle size to be $0.1~{\rm {\mu}m}$. We also tested the smallest sizes of $0.01$ and $1\,{\rm {\mu}m}$, but the results were almost unchanged.
For all of the ringed models, the intrinsic atmospheric features are much smaller than those in ring-free scenario because the surface gravity is about 6 times higher than the ring-free scenario, significantly reducing the true atmospheric scale height and thus the spectral features.

We fit all of the models described above to the observed WFC3 transmission spectrum (excluding the \textit{K2} point) by computing the mean model prediction of each spectroscopic channel and performing a least-squares fit of the band-averaged model to the spectrum. In our fits, we preserved the shape of the model by allowing the vertical offset in $\rm R_{p}/R_{\star}$ between the spectrum and model to vary while holding all other parameters fixed. The number of degrees of freedom for each model is $n$ $-$ $m$, where $n$ is the number of data points and $m$ is the number of fitted parameters. Since $n$ = 30 for the HST spectrum and $m$ = 1, the number of degrees of freedom is the same for each model.

From the fits, we quantified our model selection by computing the $\chi_{r}^{2}$ statistic. Figure \ref{fig:hip41378_trspec} shows the clear atmosphere models, the best-fitting hazy and ringed models, and a flat model. We rule out clear, low metallicity atmospheres ($\chi_{r}^{2}$=5.80 and 8.44 for the 1$\times$ and 30$\times$ solar cases, respectively). We cannot, however, distinguish between the high-metallicity (300$\times$ solar; $\chi_{r}^{2}$=1.84), hazy (soots, prod=$10^{-13}$\,$\rm g~\rm cm^{-2}~\rm s^{-1}$; $\chi_{r}^{2}$=0.97), gray opacity ringed model ($\rm R_{in}$=$1.08\rm R_{0}$, $i_{\rm ring}$=28$^{\circ}$; $\chi_{r}^{2}$=1.04) and non-gray ringed case ($\rm R_{in}$=$1.07\rm R_{0}$, $i_{\rm ring}$=28$^{\circ}$, $\rm a_{max}$=10\,$\mu$m; $\chi_{r}^{2}$=1.04) with the current observations. A flat spectrum (gray atmosphere) also matches the data well ($\chi_{r}^{2}$=1.05).  


\vspace{-0.2cm}

\section{Discussion}
\label{sec:discussion}

Based on the observed WFC3 transmission spectrum, we 
contextualize HIP\,41378\,$\rm f$   by comparing to other planets with similar masses and radii (Section~\ref{ssec:context}) and constraining the composition of putative ring particles (Section \ref{sec:rings_in_lc}). We then explore how future JWST transit observations could break the degeneracy between high-altitude hazes and circumplanetary rings (Section~\ref{ssec:mir}). We also compare the observed WFC3 transit midpoint to previous predictions and calculate the times of upcoming transits for HIP\,41378\,$\rm f$   (Section~\ref{ssec:ttvs}).  


\subsection{Placing HIP\,41378\,f in Context}
\label{ssec:context}

HIP\,41378\,$\rm f$ ($\rm R_p$ = 9.2$\pm$0.1\,$\rm R_\oplus$) is approximately the same size as Saturn ($\rm R_p$ = 9.449\,$\rm R_\oplus$) but has a much lower mass (12$\pm$3\,$\rm M_\oplus$ versus $95.16\,\rm M_\oplus$) and density ($0.09\pm0.02 \, {\rm g}\, {\rm cm}^{-3}$ versus $0.687 \,{\rm g}\, {\rm cm}^{-3}$). Although HIP\,41378\,$\rm f$   is also less dense than other exoplanets with similar radii or masses, it is not the only known low-density Saturn-sized planet. There are currently five planets with radii of $7 \rm R_\oplus < \rm R_p < 10 \rm R_\oplus$ and densities lower than $0.15 \, {\rm g}\,{\rm cm}^{-3}$: Kepler-177\,c \citep{vissapragada_et_al2020}, Kepler-51\,b, c, d \citep{ masuda2014,libby-roberts_et_al2020}, and Kepler-79\,d \citep{jontof-hutter_et_al2014, chachan_et_al2020}.  
Kepler-51\,b, Kepler-51\,d, and Kepler-79\,d have previously been observed in transmission using WFC3 and displayed (within the precision of those observations) flat, featureless transmission spectra consistent with high-altitude aerosols \citep{libby-roberts_et_al2020, chachan_et_al2020}. HIP\,41378\,$\rm f$ displays a similarly flat spectrum (see Figure~\ref{fig:hip41378_trspec}), suggesting that flat spectra may be a hallmark of temperate, ultra-low density planets. 

Several theories have been proposed to explain the extremely low density planets discovered to-date. The large radii of the more highly irradiated planets could be attributed to ohmic dissipation \citep{pu+valencia2017} or obliquity tides \citep{millholland2019}, but these mechanisms are not expected to be significant heating sources for wide-orbit, cooler planets like HIP\,41378\,$\rm f$.  A large ($>$10\%) gas mass fraction can naturally lead to an inflated radius, though acquiring and maintaining such an envelope may require formation near the water ice line and inward migration \citep{lee+chiang2016} as well as a low rate of atmospheric loss. High-altitude aerosols can reduce the gas mass fraction needed to produce the observed radii and explain the flat transmission spectra \citep[e.g.,][]{gao+zhang2020,Ohno2021}. At the low equilibrium temperature of HIP\,41378\,$\rm f$   ($\rm T_{eq}$=294~K), methane is the dominant carbon carrier in a solar metallicity atmosphere and therefore organic hazes are likely. Sulfur hazes produced from H$_{2}$S photochemistry is also possible (\citealt{Zahnle2016,gao_et_al2017}). 

Alternatively, the planets themselves could have higher densities but are surrounded by extended ring systems at an orientation that inflates their transit depths \citep{piro+vissapragada2020, akinsanmi_et_al2020}.
All of the observed low density exoplanets, including HIP\,41378\,$\rm f$, are close enough to their host stars that any ring systems would be warmer than the water ice sublimation temperature ($T_{\rm sub} \approx 170$\,K) and must therefore be composed of rocky particles rather than icy particles \citep{gaudi_et_al2003, piro+vissapragada2020} with densities of $2-5  \, {\rm g}\,{\rm cm}^{-3}$, depending on the specific particle composition and porosity. The observed transit depths of Kepler-51\,b, c, d and Kepler-79\,d, for example, are so large that they can be explained by rocky rings only if the ring material is extremely porous -- which might be possible if the ring material is particularly weak \citep{hedman2015} or similar in composition to low-density asteroids \citep{carry2012}. It is thus more challenging to explain these planets with rocky rings.

\begin{figure}
    \centering
    \includegraphics[trim={2cm 1cm 1cm 0},scale=0.50]{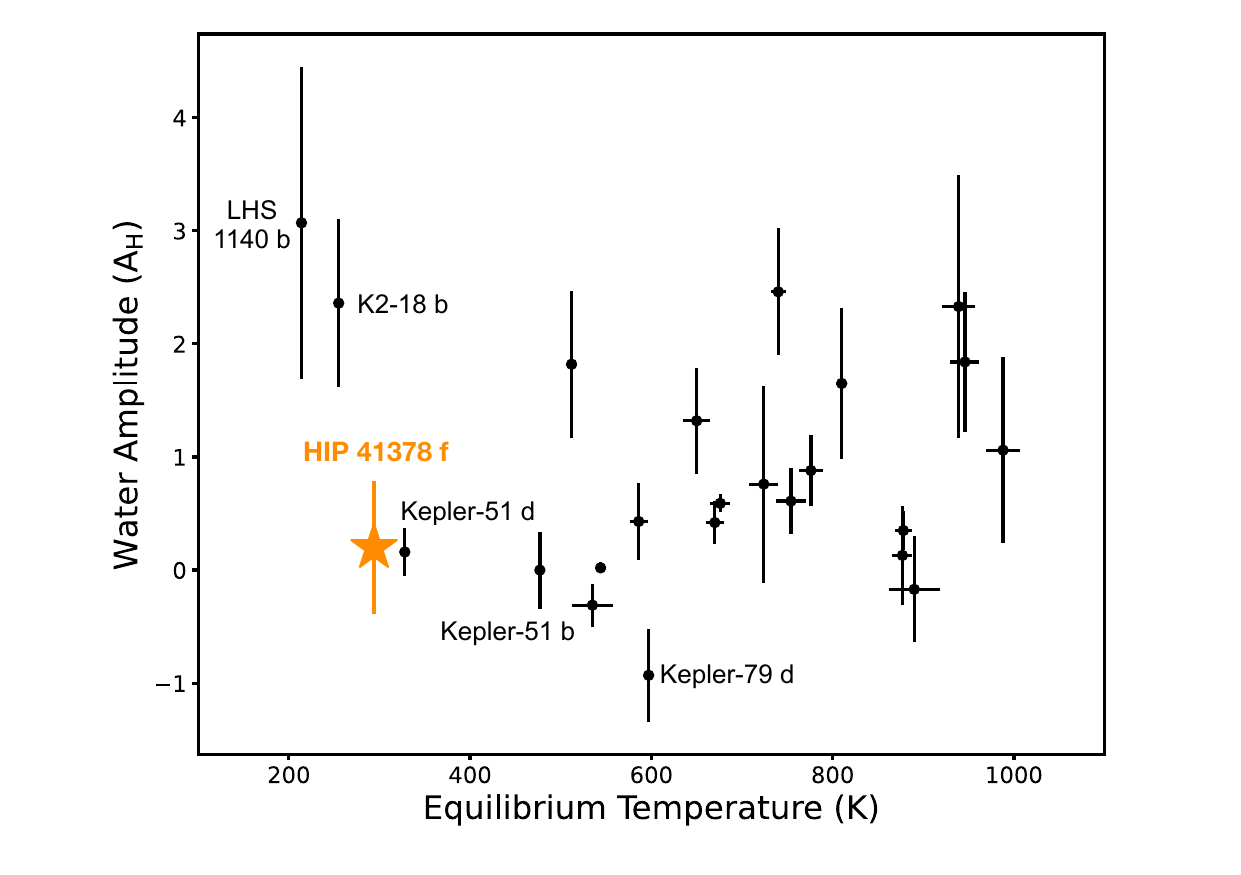}
    \caption{Equilibrium temperature versus 1.4 $\mu$m H$_{2}$O feature amplitude ($\rm A_{H}$) for the cooler ($\rm T_{eq}<$~1000\,K) planets sample from \citealt{Dymont+21} (points) and HIP\,41378\,$\rm f$ (orange star). There is no clear trend in $\rm A_{H}$ with planetary temperature.} 
    \label{fig:Teq_Ah}
\end{figure}

An emerging trend in the haziness of cooler planets (see \citealt{Crossfield17,libby-roberts_et_al2020,Yu21,Dymont+21}) hints that planets with \mbox{$\rm T_{eq}$ $<$ 300\,K} may have clear atmospheres, as possibly shown by \mbox{K2-18\,b} ($\rm T_{eq}$=282\,K; \citealt{Benneke19,Tsiaras19}) and LHS~1140\,b ($\rm T_{eq}$=229\,K; \citealt{Edwards20}). Following \citet{Dymont+21}, we compute the 1.4 $\mu$m H$_{2}$O feature amplitude ($\rm A_{H}$) and add HIP\,41378\,$\rm f$   to the sample of cooler ($\rm T_{eq}$ $<$ 1000\,K) planets with measured WFC3 transmission spectra (Figure \ref{fig:Teq_Ah}). We do not find a potentially linear \citep{Crossfield17,libby-roberts_et_al2020} or quadratic \citep{Yu21} trend in $\rm A_{H}$ with planetary equilibrium temperature, as previously suggested in the literature.  

This larger sample reiterates that cloudiness/haziness in exoplanet atmospheres is governed by complex chemical and physical processes that are controlled by multiple parameters. Despite their comparable irradiation levels, for example, the transmission spectrum of \mbox{K2-18\,b} displays an atmospheric signature of H$_{2}$O whereas the spectrum of HIP\,41378\,$\rm f$ is essentially featureless. The different emergent spectra for these two planets with similar equilibrium temperatures may be due to their distinct bulk properties. Conversely, the observed atmospheric signal for K2-18\,b may be caused by stellar surface inhomogeneities \citep{Barclay21}.  

\subsection{Modeling Ring Compositions}
\label{sec:rings_in_lc}

We use the WFC3 white light curve to constrain the composition of putative ring particles for HIP\,41378\,$\rm f$. We infer the ring properties following the framework of \citet{akinsanmi_et_al2020}, which found that the observed transit depth of HIP\,41378\,$\rm f$   could be explained by a ring system extending from $1.05$-$2.59\rm R_p$ and inclined by $\sim$25$^\circ$. In this scenario, the underlying planet would have a higher density ($1.2 \pm 0.4  \, {\rm g}\,{\rm cm}^{-3}$) and a smaller radius ($3.7^{+0.3}_{-0.2}\,R_\oplus$), while the ring particles would possess a density of $\rho_r = 1.08\pm0.30 \, {\rm g}\,{\rm cm}^{-3}$ -- lower than expected for rocky materials but comparable to the densities of porous materials comprising some asteroids \citep{carry2012}. 

Despite the lower signal-to-noise and time sampling of the WFC3 observations, we performed a joint fit to the HST white light curve and \textit{K2} (Campaigns 5 and 18) light curves, allowing for different underlying planetary radii in the different bandpasses. Our ringed model fit (constrained mostly by the \textit{K2} data) provides a ring density estimate of 1.07$\pm$0.27\,g cm$^{-3}$, consistent with results in \citet{akinsanmi_et_al2020}. The ringed model fit\footnote{The ringed model light curve fit is available \href{https://doi.org/10.6084/m9.figshare.17374061}{here} in the online supplementary material.} suggests an underlying planetary radius of $3.7^{+0.3}_{-0.2}\,R_\oplus$ for the \textit{K2} data and $3.9^{+1.2}_{-0.4}\,R_\oplus$ for HST.

\subsection{Distinguishing Between Rings, Hazes, and High Atmospheric Metallicity}
\label{ssec:mir}

The featureless near-infrared transmission spectrum of HIP\,41378\,$\rm f$ (Figure \ref{fig:hip41378_trspec}) might suggest the presence of circumplanetary rings -- although high-altitude hazes, a combination of rings and hazes, or a high mean molecular weight atmosphere could also explain the lack of spectral features. Rings composed of large ($>$10 $\mu$m) particles would result in a fairly flat spectrum with weak spectral features in the limit that it dominates over any signal from the planet's atmosphere \citep{Ohno2021,Ohno&Fortney22}. In contrast, hazes can flatten spectra fairly easily, as has been observed for other planets (e.g., \citealt{Kreidberg14,libby-roberts_et_al2020}). 

An enticing prospect for breaking the degeneracy between rings, metallicity, and aerosols is to obtain transmission spectra at near- and mid-infrared wavelengths. As shown by \citet{Ohno2021}, a super-puff with a hazy atmosphere would be expected to have a strongly sloped transmission spectrum in which the transit depth is much larger at bluer wavelengths than at redder wavelengths. This effect occurs because of the anticipated small size ($<$1 $\mu$m) of lofted dust particles. Conversely, planetary rings are likely composed of significantly larger particles, leading to less variation in transit depth with wavelength. 

Using {\tt PandExo} \citep{batalha_et_al2017}, we simulated JWST observations\footnote{The simulated JWST observations from {\tt PandExo} are included \href{https://doi.org/10.6084/m9.figshare.17960111}{here} in the online supplementary material.} of a single transit with MIRI LRS ($\sim$5-12\,$\mu$m), NIRSpec Prism ($\sim$0.5-5.5\,$\mu$m), NIRISS SOSS order 1 ($\sim$0.6-3\,$\mu$m), and NIRCam f322 ($\sim$2.4-4\,$\mu$m). At a resolution $R = 100$, we find that we can measure the transit depth for the high-metallicity clear, hazy, and non-gray rings scenarios to precisions of 85-100\,ppm with MIRI, 325-420\,ppm with NIRSpec Prism, 230-310\,ppm with NIRISS, and 220-280\,ppm for NIRCam f322. Observations with MIRI, NIRSpec Prism, NIRISS SOSS, and NIRCam f322 would be able to distinguish between the hazy and clear high-metallicity cases at 3.6-$\sigma$, 5.1-$\sigma$, 4.8-$\sigma$, and 3.4-$\sigma$, between hazy and ringed cases at 5.4-$\sigma$, 6.6-$\sigma$, 6.0-$\sigma$, and 5.1-$\sigma$, or between ringed and clear high-metallicity cases at 1.6-$\sigma$, 1.5-$\sigma$, 1.2-$\sigma$, and 1.7-$\sigma$, respectively. As shown in Figure~\ref{fig:hip41378_trspec}, the models differ in both transit depth and the slope across the near- and mid-infrared. Given the intrinsic challenge of measuring absolute transit depths, the broader wavelength coverage of MIRI and NIRISS SOSS is advantageous because of the increased ability to measure trends in transit depth with wavelength. The ability of JWST to observe continuously for the full transit also provides the opportunity to reveal subtle ring-induced deviations near ingress and egress \citep{akinsanmi_et_al2020}.
\vspace{-0.44cm}

\subsection{Future Transits}
\label{ssec:ttvs}

\begin{figure}
    \centering
    \includegraphics[trim={0 0.2cm 0 0},scale=0.56]{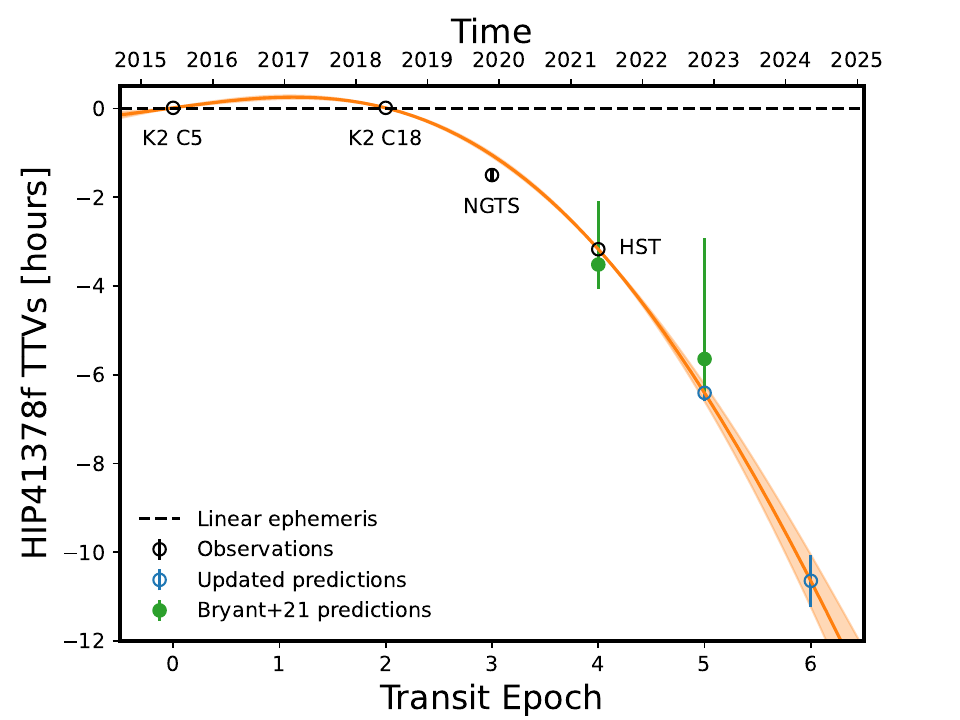}
    \caption{Transit times versus transit epoch for HIP\,41378\,$\rm f$ for observed transits from previous \textit{K2} and NGTS analyses (\citealt{vanderburg_et_al2016,becker_et_al2019, bryant_et_al2021}) compared to the current HST analysis (open black circles). We compare the transit times predicted by \citealt{bryant_et_al2021} (green circles) with our new transit predictions based on the HST transit mid-point (open blue circles). The dashed black line marks the linear ephemeris calculated using the period and transit mid-point from \citet{santerne_et_al2019}, along with the median TTV signal (orange line) and 1-$\sigma$ uncertainty (shaded region).}
    \label{fig:ttvs}
\end{figure}

To update the prediction of future transits, we reproduced the TTV analysis described in \citet{bryant_et_al2021}\footnote{\citet{bryant_et_al2021} predicted a transit center of BJD 2459355.087 with a 68\% confidence range of BJD 2459355.064 - 2459355.118 (1.3 hours long) and a 95\% confidence range of BJD 2459355.020 - 2459355.205 (4.4 hours long).}, based on the \citet{lithwick_et_al2012} formalism on the four epochs observed so far (two \textit{K2} epochs, one NGTS, and the HST transit presented in this manuscript). We assumed the timing variations of HIP\,41378\,$\rm f$ are dominated by the 2:3 resonance with HIP\,41378\,$\rm e$. As in the aforementioned study, the interaction with the very low-mass HIP\,41378\,$\rm d$ is expected to be negligible. We used {\tt emcee} to explore the posterior distribution with 40 walkers of 200,000 steps after a burn-in of 100,000 iterations. Priors were defined following the results of \citet{santerne_et_al2019}. We predict that the next transits of HIP\,41378\,$\rm f$ should occur on $\rm T_{C}$ = BJD 2459897.046  $\pm$ 0.008 (mid-transit on 2022-11-13 at 13:06:28.30 UT) and $\rm T_{C}$ = BJD 2460438.95 $\pm$ 0.02 (mid-transit on 2024-05-08 at 10:47:02.33 UT).  

As displayed in Figure~\ref{fig:ttvs}, the measured transit time is 21 minutes later than the value predicted by \citet{bryant_et_al2021} but is fully compatible with that prediction within 68.3\%. We did not detect a transit of HIP\,41378\,$\rm e$ in our HST data, which is unsurprising given the short duration of our observations compared to the length of the transit window for the planet. Given the long period of HIP\,41378\,$\rm f$, there are only a few opportunities to observe its transit during JWST's lifetime. No JWST observations are currently planned for future transits of this target, although these rare events present a unique opportunity to characterize the atmospheric properties of a cool, low mass giant planet.


\section{Conclusions}
\label{sec:conclusions}

Using HST/WFC3, we observed a transit of the low-mass, long-period temperate giant planet HIP\,41378\,$\rm f$   to measure its near-infrared transmission spectrum. Based on these measurements, our key results about the atmospheric properties of this planet and opportunities for future observations can be summarized as follows:

\begin{itemize}
    \item The transmission spectrum is featureless between $1.1-1.7\,\mu{\rm m}$, with no evidence of gaseous molecular features. Based on comparisons to 1D radiative-convective forward models, we rule out clear low metallicity atmospheres, but cannot distinguish between high metallicities, high-altitude hazes, and circumplanetary rings with the current observations. 
    \item In the context of other cooler, low density exoplanets, HIP\,41378\,$\rm f$'s featureless spectrum suggests that flat spectra are possibly a population property of ultra-low density planets. This planet also complicates the picture of cloudiness versus temperature.  
    \item Future JWST observations (e.g., MIRI, NIRSpec, NIRISS, NIRCam) can distinguish at $>$1-$\sigma$ confidence between the super-puff scenario in which HIP\,41378\,$\rm f$   is a low-density planet shrouded in a high-altitude aerosol layer, the ringed scenario in which the planet itself is much smaller than expected from the observed optical and near-infrared transit depths, and a clear high mean molecular weight atmosphere scenario. 
    \item \vspace{-10cm} We predict the next transits of HIP\,41378\,$\rm f$ to occur at BJD = 2459897.046 $\pm$ 0.008 and BJD = 2460438.95 $\pm$ 0.02. These upcoming transits provide a rare opportunity to observe the atmospheric properties of a low mass, temperate gas giant planet with JWST, thereby expanding our efforts for comparative exoplanetology.   
\end{itemize}

With the current HST observations, it is also possible to place constraints on the potential presence of exomoons. A $1.5 \, \rm R_\oplus$ moon would produce a 115~ppm transit (comparable to the precision we achieve in each spectrophotometric channel). Although a lunar transit would cause a noticeable deviation in the light curve, the moon would have to be precisely aligned and at a favorable orbital phase. A moon detection is therefore unlikely, but we will discuss the limits from this serendipitous search in a follow-up paper.

\begin{acknowledgments}
We thank the anonymous referee for their insightful comments. We appreciate the painstaking work of the HST technical staff including Patrica Royle and Nikolay Nikolov in scheduling this long sequence of observations. This paper makes use of observations from the NASA/ESA Hubble Space Telescope, obtained at the Space Telescope Science Institute, which is operated by the Association of Universities for Research in Astronomy, Inc., under NASA contract NAS 5-26555. These observations are associated with HST GO program 16267 (PI: Dressing) and the analysis was supported by grant HST-GO-16267.
M.K.A. is grateful to Johanna Teske and Anjali Piette for useful discussions. C.D.D. gratefully acknowledges additional support from the David \& Lucile Packard Foundation (grant number 2019-69648) and helpful conversations with Christina Hedges. K.O. was supported by JSPS Overseas Research Fellowship. N.S., S.B., and B.A. acknowledge the support by FCT (Funda\c{c}\~ao para a Ci\^encia e a Tecnologia) through national funds and by FEDER through COMPETE2020 - Programa Operacional Competitividade e Internacionaliza\c{c}\~ao by these grants: UID/FIS/04434/2019; UIDB/04434/2020; UIDP/04434/2020;
PTDC/FIS-AST/32113/2017 \& POCI-01-0145-FEDER-032113; PTDC/FISAST
/28953/2017 \& POCI-01-0145-FEDER-028953. V.A. acknowledges the support from FCT through Investigador contract nr. IF/00650/2015/CP1273/CT0001. J.L-B. acknowledges financial support received from ``la Caixa" Foundation (ID 100010434) and from the European Unions Horizon 2020 research and innovation programme under the Marie Slodowska-Curie grant agreement No 847648, with fellowship code LCF/BQ/PI20/11760023. This research has also been partly funded by the Spanish State Research Agency (AEI) Projects No.ESP2017-87676-C5-1-R and No. MDM-2017-0737 Unidad de Excelencia ``Mar\'ia de Maeztu"- Centro de Astrobiolog\'ia (INTA-CSIC). 
\end{acknowledgments}

\vspace{5mm}
\facilities{HST(WFC3)}

\bibliography{hip41378}{}
\bibliographystyle{aasjournal}

\end{document}